\documentclass{article}
\usepackage{spconf}
\usepackage{amsmath}
\usepackage[utf8]{inputenc} 
\usepackage[T1]{fontenc}    
\usepackage{hyperref}       
\usepackage{url}            
\usepackage{booktabs}       
\usepackage{amsfonts}       
\usepackage{nicefrac}       
\usepackage{microtype}      
\usepackage{xcolor}         
\usepackage{float}
\usepackage[thinlines]{easytable}
\usepackage{subfig}
\usepackage[export]{adjustbox}
\usepackage{graphicx}
\usepackage{pdflscape}
\usepackage{breqn}

\title{Higher-order Organization in the Human Brain \\ from Matrix-Based Rényi's Entropy}
\name
  {Qiang Li$^{1,5}$\sthanks{Corresponding author: qiang.li@uv.es}, Student Member, IEEE, Shujian Yu$^2$\sthanks{Corresponding author: s.yu3@vu.nl}, Kristoffer H Madsen$^{3,4}$, Vince D Calhoun$^5$, Armin Iraji$^5$}
\address{
$^1$Image Processing Laboratory, University of Valencia, Spain \leavevmode\\   
$^2$Department of Computer Science, Vrije Universiteit Amsterdam, The Netherlands \leavevmode\\
$^3$Department of Applied Mathematics and Computer Science, Technical University of Denmark, Denmark \leavevmode\\
$^4$Danish Research Centre for Magnetic Resonance, Copenhagen University Hospital, Denmark\leavevmode\\
$^5$Tri-Institutional Center for Translational Research in Neuroimaging and Data Science (TReNDS), USA
}

\begin{document}
\maketitle
\begin{abstract}
Pairwise metrics are often employed to estimate statistical dependencies between brain regions, however they do not capture higher-order information interactions. It is critical to explore higher-order interactions that go beyond paired brain areas in order to better understand information processing in the human brain. To address this problem, we applied multivariate mutual information, specifically, \emph{Total Correlation} and \emph{Dual Total Correlation} to reveal higher-order information in the brain. In this paper, we estimate these metrics using matrix-based \emph{Rényi's entropy}, which offers a direct and easily interpretable approach that is not limited by direct assumptions about probability distribution functions of multivariate time series. We applied these metrics to resting-state fMRI data in order to examine higher-order interactions in the brain. Our results showed that the higher-order information interactions captured increase gradually as the interaction order increases. Furthermore, we observed a gradual increase in the correlation between the \emph{Total Correlation} and \emph{Dual Total Correlation} as the interaction order increased. In addition, the significance of \textit{Dual Total Correlation} values compared to \textit{Total Correlation} values also indicate that the human brain exhibits synergy dominance during the resting state.
\end{abstract}

\begin{keywords}
Higher-order Interactions, Total Correlation, Dual Total Correlation, Rényi's Entropy, Resting State
\end{keywords}

\section{Introduction}
\label{sec:intro}
The brain is densely interconnected to facilitate efficient information transfer. The brain can be modeled as a fractal, chaotic, and complex system~\cite{LiebovitchLarryS1998Facs}, and our understanding of the brain requires studying how higher-order information is distributed via networks at different levels (e.g., microscopic versus macroscopic, where we are concerned with the latter in this study). 

Higher-order information enables intricate interactions among brain regions, these interactions encompass not only paired connections across brain regions, but also connections between three or more regions, as well as more complex and dynamic activity patterns including synchronization and phase relationships~\cite{hancock2022metastability}. Furthermore, comprehending higher-order functional interactions is essential for understanding how the brain performs complicated activities and identifying biomarkers that can lead to improved diagnosis and treatment of neurological and psychiatric disorders~\cite{li2022functionalen,herzog2022genuine}.

There are two general categories to investigate higher-order information interactions in the brain. The first category estimates higher-order interaction among brain networks using a hypergraph representation~\cite{battiston2020networks,santoro2023higher}. The second category is information-theoretic approaches and uses the information shared across regions to estimates estimate high-order interactions ~\cite{li2022functionalen,rosas2019quantifying,li2022functional,li2022functionalnn}. Our focus in this paper is primarily on the latter approach, which aims to reveal higher-order information interactions in the human brain via information theory. For this purpose, we will use the two most commonly used metrics: (1) \textit{Total Correlation (DC)}~\cite{watanabe1960information} (also known as \textit{multi-information}~\cite{studeny1998multiinformation} or \textit{redundancy}~\cite{te1980multiple}) and (2) \textit{Dual Total Correlation (DTC)}~\cite{Han78} (also known as \textit{binding information}~\cite{vijayaraghavan2017anatomy} or \textit{excess entropy}~\cite{olbrich2008should}).

There are various ways to estimate \textit{TC} and \textit{DTC}. One popular method is the Gaussian copula transform~\cite{ince2017statistical}, which involves transforming non-Gaussian distributed multivariate time series into a Gaussian distribution based on a \textit{copula transform} and then estimating the above metrics from their covariance~\cite{ince2017statistical}. However, in practice this method may face a range of challenges, for example, the power of these methods relies on the fact that the underlying fMRI BOLD signals can be well modeled by multivariate Gaussian distributions and thus allow efficient, easy, and analytically tractable estimation~\cite{hlinka2011functional}. However, the use of this kind of processing ignores non-Gaussian distributed signals and does not provide a universal way to estimate them. 

In this study, we used a more comprehensive approach, a matrix-based \emph{Rényi's entropy} for estimating higher-order information interactions and quantifying information organization in the human brain.  Matrix-based \textit{Renyi's entropy}~\cite{yu2019multivariate,yu2021measuring} estimates information in data without the need to \textcolor{black}{know} the underlying data distribution. This is achieved by utilizing the eigenspectrum of symmetric positive semi-definite matrices in a reproducing kernel Hilbert space. In our study, we investigate how higher-order information interactions can be captured by considering interaction orders beyond pairwise brain regions. Our specific interest lies in examining how the differences between \textit{TC} and \textit{DTC} vary as the interaction order increases, and how these metrics relate to the role of higher-order information in resting-state brain functions. 

\section{Methodology}
\label{sec:metho}
\subsection{\textit{Renyi’s $\alpha$ entropy functional}}

In information theory, a natural extension of the well-known \textit{Shannon’s entropy}~\cite{shannon1948mathematical} is the \textit{Renyi’s $\alpha$ entropy}~\cite{renyi1961measures}. 
For a random variable $X$ with probability density function $p(x)$ in a finite set $\mathcal{X}$, the $\alpha$ entropy is defined as:
\begin{equation}
    \mathbf{H}_\alpha(X)=\frac{1}{1-\alpha} \log \left(\int_{\mathcal{X}} p^{\alpha}(x) dx \right),
\end{equation}
with $\alpha \neq 1$ and $\alpha \ge 0$. In the limiting case where $\alpha \rightarrow 1$, it reduces to \textit{Shannon’s entropy}~\cite{cover1991information}.


{\color{black}
In practice, given $m$ realizations sampled from $p(x)$, i.e., $\{x_i\}_{i=1}^m$, Sanchez~Giraldo \emph{et al.}~\cite{giraldo2014measures} suggests that one can evaluate $\mathbf{H}_\alpha(X)$ without estimating $p(x)$. Specifically, the so-called \textit{matrix-based Renyi’s $\alpha$ entropy} is given as:
\begin{equation}
    \mathbf{H}_\alpha(X)=\frac{1}{1-\alpha} \log \left(\operatorname{tr}\left(A^\alpha\right)\right),
    \label{eq.ep}
\end{equation}
where $A \in \mathbb{R}^{m \times m}$ is a (normalized) Gram matrix with elements $A_{i j} = K_{i j}/\operatorname{tr}(K)$, $K_{i j}=\kappa\left(x_i, x_j\right)$ in which $\kappa$ stands for a positive definite and infinitely divisible kernel such as Gaussian. $\operatorname{tr}(*)$ refers to matrix trace. \textcolor{black}{As in~\cite{yu2019multivariate}, we set $\alpha=1.01$ to approximate Shannon entropy and choose a Gaussian kernel $G_\sigma$ with width $\sigma$, given by,
\begin{equation}
    G_\sigma\left(x_i, x_j\right)=\beta \exp \left(-\frac{\left\|x_i-x_j\right\|^2}{2 \sigma^2}\right),
    \label{eq.Gaukernel}
\end{equation}
where $\beta$ is a constant whose value is irrelevant because it is canceled out in the normalized Gram matrix.}

Suppose now we have $n\geq 2$ variables ($X^1,X^2,\cdots,X^n$) and a collection of $m$ samples\footnote{Throughout this paper, we use superscript to denote variable index and subscript to denote sample index. For example, $x^3_i$ refers to the $i$-th sample from the $3$rd variable.}, i.e., $\{x^1_i,x^2_i,\cdots,x^n_i\}_{i=1}^m$, the \textit{matrix-based Renyi’s $\alpha$ joint entropy} for $n$ variables can be evaluated as~\cite{yu2019multivariate}: 
\begin{dmath}
    \mathbf{H}_\alpha\left(X^1, X^2, \cdots, X^n\right)=\mathbf{H}_\alpha\left(\frac{K^1 \circ K^2 \circ \cdots \circ K^n}{\operatorname{tr}\left(K^1 \circ K^2 \circ \cdots \circ K^n\right)}\right)=
    \frac{1}{1-\alpha} \log \left(\operatorname{tr}\left(\left(\frac{K^1 \circ K^2 \circ \cdots \circ K^n}{\operatorname{tr}(K^1 \circ K^2 \circ \cdots \circ K^n)}\right)^\alpha\right)\right),
    \label{eq.muljoin}
\end{dmath}
where $\left(K^n\right)_{i j} = \kappa(x^n_i,x^n_j) \in \mathbb{R}^{m \times m}$ is a Gram matrix evaluated with $\kappa$ for the $n$-th variable. The operator $\circ$ is the Hadamard product.
}

\subsection{Bivariate dependencies from \textit{Mutual Information}}

Bivariate dependencies can be inferred from \textit{mutual information}, and if two variables are directly dependent, then the values are zeros. Given Eqs.~\ref{eq.ep} and~\ref{eq.muljoin}, the \textit{Renyi’s $\alpha$ entropy mutual information} $\mathbf{I}_a(X^1, X^2)$ between variables $X^1$ and $X^2$ in analogy of \textit{Shannon’s mutual information} is given by:
\begin{equation}
    \mathbf{I}_\alpha(X^1 ; X^2)=\mathbf{H}_\alpha(X^1)+\mathbf{H}_\alpha(X^2)-\mathbf{H}_\alpha(X^1, X^2).
\end{equation}

\textit{Mutual information} is the most widely used metric for quantifying statistical dependency between two variables~\cite{cover1991information}, \textcolor{black}{but since it omits} higher-order interactions in the system it does not capture all global information in the system.

\subsection{Inferring higher-order dependencies through \textit{Matrix-based Rényi's $\alpha$ TC} and \textit{DTC}}

The \textit{TC} and \textit{DTC} describe the dependence among $n$ variables and can be considered as a non-negative generalization of the concept of \textit{mutual information} from two parties to $n$ parties. \textcolor{black}{The definition of \textit{TC} can be denoted as follows~\cite{watanabe1960information}:}
\begin{equation}\label{eq.tc}
\begin{split}
   & \mathbf{T}\left(X^1, X^2, \cdots, X^{n}\right) = \sum_{i=1}^n \mathbf{H}\left(X^{i}\right)-\mathbf{H}\left(X^1, X^2 \cdots, X^{n}\right). 
\end{split}
\end{equation}


Analogously with \textit{TC}, \textit{DTC} can be defined as~\cite{Han78},
\begin{equation}\label{eq.dtc}
\begin{split}
    & \mathbf{D}\left(\textcolor{black}{X^1, X^2} \cdots, X^n\right)=\mathbf{H}\left(\textcolor{black}{X^1, X^2} \cdots, X^n\right)-\sum_{i=1}^n \mathbf{H}\left(X^i \mid X^{[n] \backslash i} \right) \\
    &  =\left[\sum_{i=1}^n \mathbf{H}\left(X^{[n] \backslash i}\right)\right]-(n-1) \mathbf{H}\left(\textcolor{black}{X^1, X^2}, \cdots, X^n\right),
\end{split}
\end{equation}
where $X^{[n] \backslash i}= \{X^1, X^2 \cdots, X^{i-1}, X^{i+1}, \cdots, X^n\}$, i.e., the set of all variables excluding $X^i$. From these definitions in Eqs.~\ref{eq.tc},~\ref{eq.dtc}, if all variables are independent, both \textit{TC} and \textit{DTC} will be zero. Beyond this, \textit{TC} and \textit{DTC} are closely related, 
\begin{dmath}
    \mathbf{D}\left(X^1, X^2 \cdots, X^n\right) \leq(n-1) \mathbf{T}\left(X^1, X^2 \cdots, X^n\right).
\end{dmath}

\begin{figure}[htb]
\begin{minipage}[b]{1.0\linewidth}
  \centering
  \centerline{\includegraphics[width=0.92\textwidth, height=5.5em]{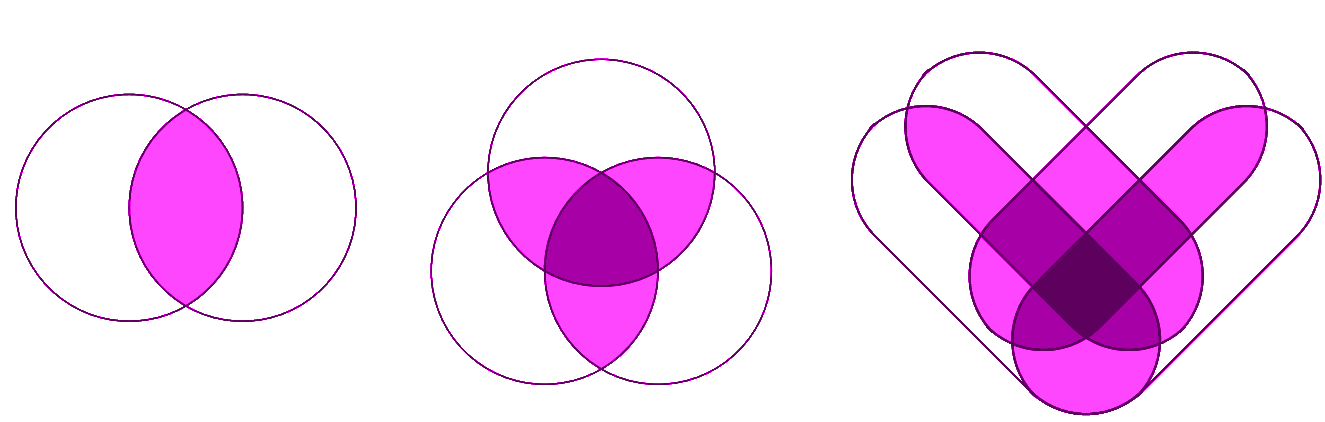}} 
  \textcolor{black}{\rule{8cm}{1mm}}
  \centerline{\includegraphics[width=0.92\textwidth, height=5.5em]{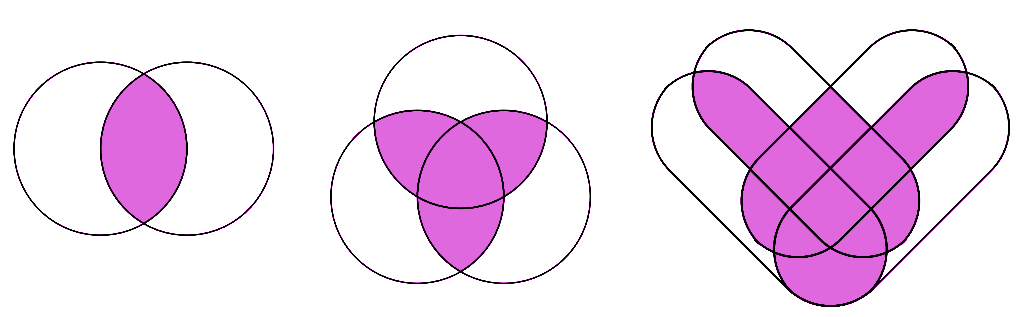}}
  \end{minipage}
\caption{\textbf{Information diagrams for \textit{TC} (top) and \textit{DTC}(bottom).} The top depicts the information interaction of 2-variable, 3-variable, and 4-variable measures from \textit{TC}, while the bottom depicts the information interaction of 2-variable, 3-variable, and 4-variable measures from \textit{DTC}. For a 3-variable case, the deep purple indicates information duplicated twice, the light purple indicates information duplicated once, and so on for a 4-variable situation.}
\label{fig:infodiag}
\end{figure}

\textcolor{black}{In cases 2-variable, 3-variable, and 4-variable, the venn diagrams (see Fig.~\ref{fig:infodiag}) provide a graphical representation of \textit{TC} and \textit{DTC}.}

In order to estimate \textit{TC} and \textit{DTC} in a practical setting only from $m$ samples $\{x^1_i,x^2_i,\cdots,x^n_i\}_{i=1}^m$, we convert Eqs.~\ref{eq.tc} and \ref{eq.dtc} to matrix-based \textit{Renyi's $\alpha$ entropy functional} based on Eqs.~\ref{eq.ep} and \ref{eq.muljoin}, which simplifies the estimation~\cite{yu2019multivariate,yu2021measuring} and enables a reformulation:
\begin{equation}
\begin{split}
& \mathbf{T}_\alpha(X^1,X^2,\cdots,X^n)=\sum_{i=1}^n \mathbf{H}_\alpha(X^i) -\mathbf{H}_\alpha(X^1,X^2,\cdots,X^n) \\
& =\left[ \sum_{i=1}^n \frac{1}{1-\alpha} \log \left(\operatorname{tr}\left(\frac{K^i}{\operatorname{tr}(K^i)}\right)^\alpha\right) \right] \\
&-\frac{1}{1-\alpha} \log \left(\operatorname{tr}\left(\left(\frac{K^1 \circ K^2 \circ \cdots \circ K^n}{\operatorname{tr}(K^1 \circ K^2 \circ \cdots \circ K^n)}\right)^\alpha\right)\right),
\end{split}
\end{equation}
\begin{multline}
    \mathbf{D}_\alpha(X^1, X^2, \cdots, X^n)=\left[\sum_{i=1}^n \mathbf{H}_\alpha\left(X^{[n] \backslash i}\right)\right] \\
    -(n-1)\mathbf{H}_\alpha(X^1,X^2,\cdots,X^n).
\end{multline}

If $\mathbf{D}_\alpha$ $>$ $\mathbf{T}_\alpha$, the system becomes synergistic and dominant, while the system becomes redundancy dominant in the opposite case~\cite{rosas2019quantifying}. 

\section{Experiments}
\label{sec:exp}

\subsection{Higher-order information in the human brain}
\subsubsection{HCP dataset}
The preprocessed time series obtained from rsfMRI data were utilized in\textcolor{black}{~\cite{Van13}}. For details regarding the scanning parameters and related preprocessing for both structural and functional MRI, refer to\textcolor{black}{~\cite{Van13}}. After preprocessing, each subject was represented by a set of 116 time series of rsfMRI signals, consisting of 1200 samples with a repetition time of 720 ms, each corresponding to a specific region of interest. To limit the computational complexity, we randomly selected 10 subjects that were not related and applied multivariate information-theoretic metrics to quantify high-order information interaction in the brain.


\subsubsection{Brain atlas and interaction order} 
\begin{figure*}[htb]
\begin{minipage}[b]{1.0\linewidth}
  \centering
  \centerline{\includegraphics[width=0.96\textwidth, height=12em]{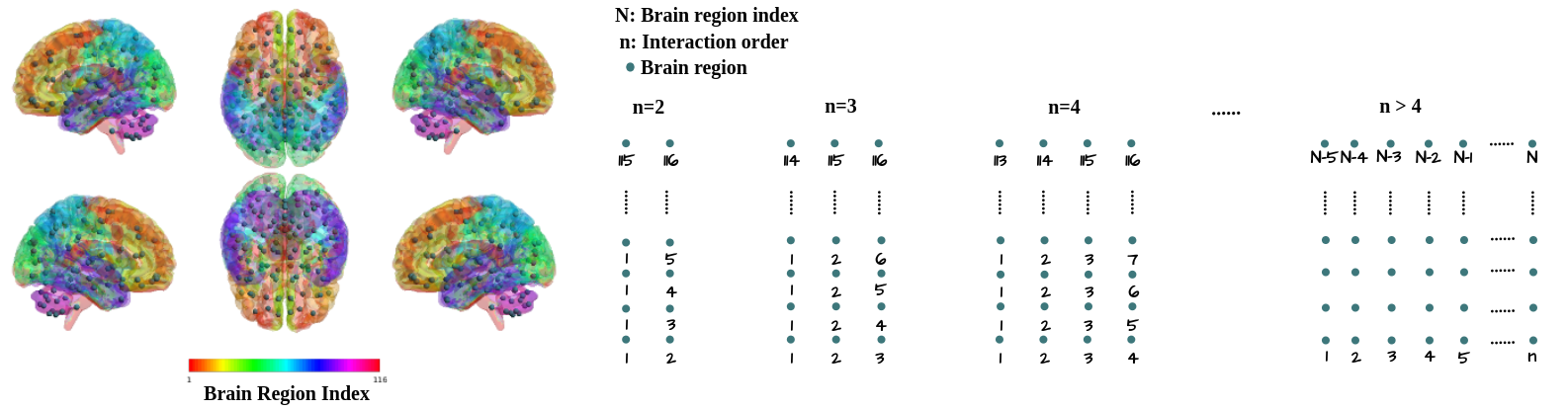}}
\end{minipage}
\caption{\textbf{Brain parcellation and possible interaction order in the brain.} The left side is the AA116 brain atlas, and the right side is the various interaction orders in the human brain. For more information, see Section 3.1.2.}
\label{fig:aal116}
\end{figure*}

The whole brain was divided into 116 brain areas using the AAL116 brain atlas~\cite{tzourio2002automated}. Considering the excessive computational cost, we limited our examination of the interaction order to only 2, 3, and 4 brain regions. \textcolor{black}{Fig.~\ref{fig:aal116} presents} the general framework, and the left panel shows the AAL116 regions (with nodes representing the 116 brain regions and their corresponding time series) mapped to \textcolor{black}{the Colin27 smoothed brain surface template~\cite{Holmes98}}. The different brain lobes are labeled on the brain surface with different colors. The right panel is a representation of multiple interaction orders in the brain, such as interaction orders 2, 3, and 4. The green node indicates brain regions, and the number, N=1, 2, $\cdots$, 116 is the index for brain regions.

\begin{figure*}[htb]
\begin{minipage}[b]{1.0\linewidth}
  \centering
  \centerline{\includegraphics[width=0.9\textwidth, height=10em]{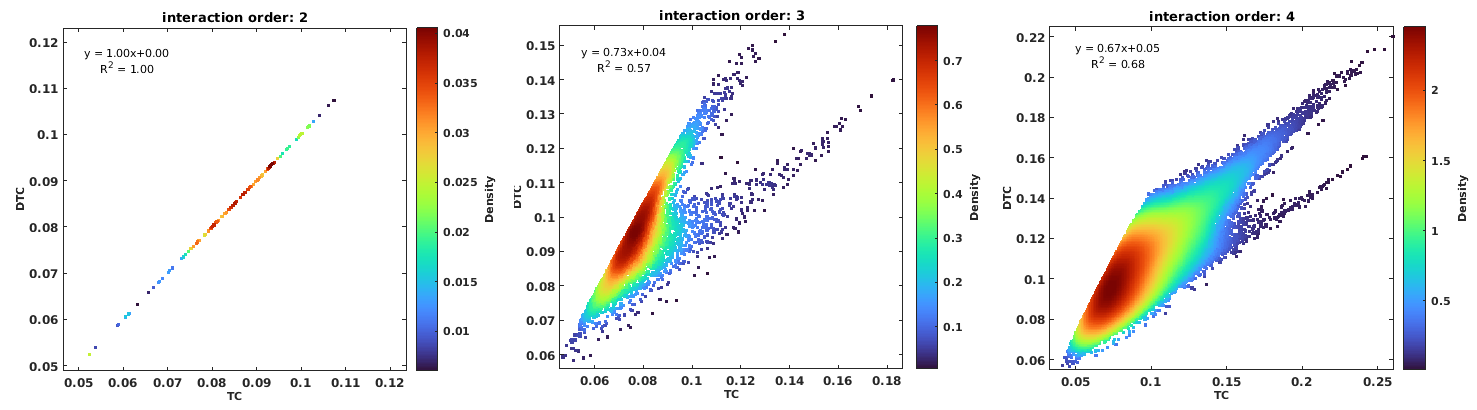}}
  \textcolor{black}{\rule{10cm}{0.6mm}}
  \centerline{\includegraphics[width=0.9\textwidth, height=9.5em]{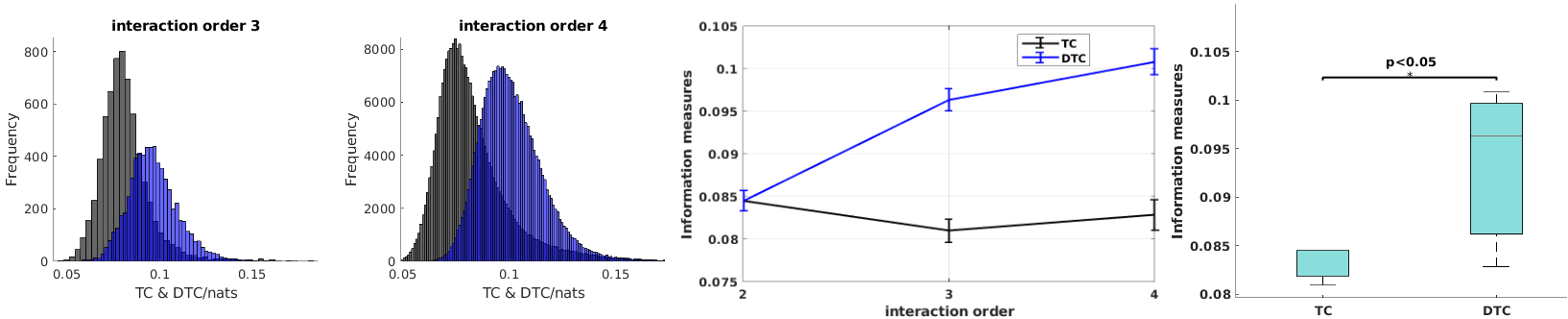}}  
\end{minipage}
\caption{\textbf{Higher-order information interaction in the brain.} The upper panels  illustrates how the brain represents higher-order information under resting state, at increasing interaction orders of 2, 3, and 4. The lower panels presented some related statistical tests. For further details, please refer to the explanation provided in the conclusion section.}
\label{fig:interOrd}
\end{figure*}

\section{CONCLUSIONS AND FUTURE WORK}
\label{sec:dis}
We present a matrix-based approach to estimate higher-order information using \textit{Rényi's entropy}, which provides a more simplified method for quantifying such information. We further quantified higher-order information using resting-state fMRI as presented in Figure~\ref{fig:interOrd}, where \textit{TC} and \textit{DTC} were utilized. It is evident from the plots and the second-order polynomial fits that \textit{TC} and \textit{DTC} are equivalent (\textcolor{black}{$R^{2}$=1.00}) because they both represent \textit{mutual information} at interaction order 2. However, with interaction orders exceeding 2 (\textcolor{black}{e.g., 3 and 4}), higher-order information is captured, and as the interaction order increases, the correlation between \textit{TC} and \textit{DTC} increases (\textcolor{black}{e.g., $R^{2}$=0.57, 0.68}), and \textit{DTC} is greater than \textit{TC} (as shown in the first row). \textcolor{black}{Furthermore, the average \textit{TC} and \textit{DTC} for 10 subjects are represented for each interaction order in the second row middle Figure~\ref{fig:interOrd}, and when we combined all interaction orders together (as shown in the second row, right panel), our finding that \textit{DTC} is more significant than \textit{TC} (p<0.05) in resting-state brain activity suggests that synergy is dominant during resting-state situations in the human brain due to default mode networks, which is consistent with previous findings~\cite{luppi2022synergistic}}.

Here, we initially limited our analysis to interaction orders of 2, 3, and 4 due to computational constraints. However, it would be intriguing to investigate higher interaction orders in future research. Secondly, from a technical perspective, \textit{TC} and \textit{DTC}, can also be used to estimate redundancy and synergy, respectively~\cite{rosas2019quantifying}. It will be an important next goal to extend this research to investigate these aspects further. Thirdly, in terms of practical applications, it is promising to use these information-theoretical metrics to quantify changes in higher-order information interaction in major brain disorders and to extend the investigations to multiple modalities. \textcolor{black}{Fourthly, using more large and task-related fMRI datasets to test the above conclusion would be very interesting in future work, and even more, considering the computation complexity for higher-order brain region interaction, we should consider using independent component networks instead of regions of interest.}

\bibliographystyle{IEEE}
\bibliography{strings}
\end{document}